\documentclass{LMCS}

\def\doi{8 (2:17) 2012}
\lmcsheading%
{\doi}
{1--18}
{}
{}
{Jun.~16, 2011}
{Jun.~28, 2012}
{}

\usepackage{latexsym}
\usepackage{amssymb}
\usepackage{amsmath}
\usepackage{graphicx}
\usepackage{gastex}
\usepackage{color}
\usepackage{wrapfig}
\usepackage{rotating}
\usepackage{enumerate,hyperref}

\def\high{\mathit{high}}
\def\medium{\mathit{medium}}
\def\low{\mathit{low}}

\def\B{\mathcal{B}}
\def\C{\mathcal{C}}
\def\F{\mathcal{F}}
\def\P{\mathbf{P}}
\def\Pr{\mathrm{Pr}}
\def\I{\mathbf{I}}
\def\R{\mathbf{R}}
\def\phi{\varphi}
\def\Sat{\mathit{Sat}}
\def\Prob{\mathcal{P}}
\def\mywedge{{\wedge}}

\newcommand{\partialto}{\mathrel{\makebox[0pt][l]{\hspace*{0.35em}\raisebox{0.1ex}[0ex]{$\shortmid$}}\mathord{\rightarrow}}}

\newcommand{\ps}[2]{#1 #2}
\newcommand{\psd}[2]{\ps{#1}{#2} \ldots}

\newcommand{\dtmc}{\mathcal{D}=(S,\mathbf{P},L)}
\newcommand{\ctmc}{\mathcal{C}=(S,\R,L)}
\newcommand{\infpaths}{\mathit{Path}}
\newcommand{\until}[1]{\mathrel{U_{#1}}}
\newcommand{\diam}[1]{\Diamond_{#1}}

\newcommand{\dnj}[1]{{\color{red}\textbf{[\!\![\!\![~DJ: #1 ---DJ.~]\!\!]\!\!]}}}
\newcommand{\lz}[1]{\color{blue}\textbf{[\!\![\!\![~LZ: #1 ---LZ.~]\!\!]\!\!]}\color{black}}

\begin{document}
\title[Efficient CSL Model Checking Using Stratification]{Efficient CSL Model Checking Using Stratification\rsuper*}

\author[L.~Zhang]{Lijun Zhang\rsuper a}
\address{{\lsuper{a,c}}Technical University of Denmark,
	DTU Informatics, Denmark}
\email{\{zhang,nielson\}@imm.dtu.dk}

\author[D.~N.~Jansen]{David N.~Jansen\rsuper b}
\address{{\lsuper b}Radboud Universiteit, Model-based System Design,
	Nijmegen, The Netherlands}
\email{dnjansen@cs.ru.nl}

\author[F.~Nielson]{Flemming Nielson\rsuper c}
\address{\vskip-6 pt}

\author[H.~Hermanns]{Holger Hermanns\rsuper d}
\address{{\lsuper d}Saarland University, Computer Science,
	Saarbr\"ucken, Germany}
\email{hermanns@cs.uni-saarland.de}


\titlecomment{{\lsuper*}A preliminary version of the paper has appeared in~\cite{icalp}.}

\begin{abstract}
  For continuous-time Markov chains, the model-checking problem with
  respect to continuous-time stochastic logic (CSL) has been
  introduced and shown to be decidable by Aziz, Sanwal, Sin\-ghal and
  Brayton in 1996 \cite{AzizSSB96,AzizSSB00}.
	Their proof can be turned into an approximation algorithm
	with worse than exponential complexity.
	In 2000, Baier, Haverkort, Hermanns and Katoen \cite{BaierHHK00,hermanns_ctmc} presented an efficient polynomial-time approximation algorithm
	for the sublogic in which only binary until is allowed.
	In this paper, we propose such an efficient polynomial-time approximation algorithm
	for full CSL.

  The key to our method is the notion of \emph{stratified CTMCs}
  with respect to the CSL property to be checked.
	On a stratified CTMC, the probability to satisfy a CSL path formula
	can be approximated by a transient analysis
	in polynomial time (using uniformization).
	We present a measure-preserving, linear-time and -space transformation
	of any CTMC into an equivalent, stratified one.
	This makes the present work the centerpiece of a
  broadly applicable full CSL model checker.

	Recently, the decision algorithm by Aziz \emph{et al.\@}
	was shown to work only for stratified CTMCs.
	As an additional contribution,
	our measure-preserving transformation
  can be used to ensure the decidability for general CTMCs.
\end{abstract}

\maketitle

\section{Introduction}
Continuous-time Markov chains (CTMC) play an important role in
performance evaluation of networked, distributed, and 
biological systems. The concept of formal verification for CTMCs was
introduced by Aziz, Sanwal, Sin\-ghal and Brayton in 1996~\cite{AzizSSB96,AzizSSB00}. Their seminal paper defined
continuous-time stochastic logic (CSL)
to specify properties over CTMCs. It showed that the model
checking problem for CTMCs, which asks whether the
CTMC satisfies a given CSL property, is decidable,
using algebraic and transcendental number theory.
	Their proof is constructive,
	so it can be turned into an approximation procedure for the relevant probabilities.
	However, its complexity may be worse than exponential in the size of the formula.

The characteristic construct of CSL is a probabilistic formula
of the form $\Prob_{<p}(\phi)$, where $p\in[0,1]$. Here $\phi$ is a path
formula; more concretely, it is a \emph{multiple until} formula
$f_1\until{I_1}f_2\until{I_2}\ldots \until{I_{k-1}} f_k$
where $k\ge 2$.
The formula $\Prob_{<p}(\phi)$ expresses a constraint on the probability
to reach an $f_k$-state
by passing only through (zero or more)
$f_1$-, $f_2$-,\break \ldots, $f_{k-1}$-states
in the given order
(together with a timing constraint
indicated by the intervals $I_1, \ldots, I_{k-1}$).
The key to solve the model checking problem is
to approximate this probability $\Pr_s(\phi)$
closely enough to decide whether it is $<p$.
The decision procedure in~\cite{AzizSSB00} first decomposes the formula
into (up to) $(k-1)^{k-1}$ many subformulas with suitable timing constraints.
For each subformula, it then exploits properties of
algebraic and transcendental numbers,
but the corresponding algorithm is unfortunately impractical.
In 2000, Baier \emph{et al.}~\cite{BaierHHK00,hermanns_ctmc} presented an \emph{approximate model
  checking algorithm} for the case $k=2$. This algorithm is
based on transient probability analysis for CTMCs. More precisely, it
was shown that $\Pr_s(\phi)$ can be approximated, up to an a priori given
precision $\varepsilon$, by a sum of transient probabilities in the
CTMCs. Their algorithm then led to further development of
approximation algorithms for infinite CTMCs \cite{HahnHWZ09,HenzingerMW09} 
and abstraction techniques \cite{KatoenKLW07}. 
More importantly, several
tools support approximate model checking, including PRISM~\cite{KwiatkowskaNP11} and MRMC~\cite{KatoenZHHJ09}.

Effective model checking of full CSL with multiple
until formulas ($k > 2$) is an open problem. This problem is
gaining importance e.\,g.\@ in the field of system biology, where
one is interested in oscillatory behavior of
CTMCs~\cite{ballarini,spieler}. 
More precisely, if one intends
to quantify the probability mass oscillating between high, medium and
low concentrations (or numbers) of some species, a formula like $\Prob_{>0.2}(\high
\until{I_1} \medium \until{I_2}\low \until{I_3} \medium \until{I_4}
\high)$ is needed, but this is not at hand with the current state of
the art.  
In CTL, multiple until formulas like $\forall (\high \until{} \medium \until{} \low \until{} \medium \until{} \high)$ do not increase expressivity
because they are equivalent to something like $\forall (\high \until{} \forall (\medium \until{} \forall (\ldots \until{} \high)))$.

In this paper we propose an approximate algorithm for checking CSL
with multiple until formulas. We introduce a subclass of
\emph{stratified CTMCs,} on which the approximation of $\Pr_s(\phi)$
can be obtained by efficient transient analysis. Briefly, a CTMC is
stratified with respect to $\phi=f_1\until{I_1}f_2\until{I_2}\ldots
f_k$, if the transitions of the CTMC respect the order given
by the $f_i$. This specific order makes it possible to express
$\Pr_s(\phi)$ recursively:
more precisely, it is the product of a transient vector and $\Pr_{s'}(\phi')$,
where $\phi'$ is a kind of suffix subformula of $\phi$.
Stratified CTMCs are the key element for our analysis:
in a stratified CTMC, the problem reduces to a transient analysis, for which
efficient implementations using \emph{uniformization}~\cite{Grassmann91} exist.
 Thus, we extend the well-known result~\cite{hermanns_ctmc}
for the case of binary until to multiple until formulas.

For a general CTMC, we present a measure-preserving transformation to a
stratified CTMC.
Our reduction is described
using a \emph{deterministic finite automaton} (DFA)
over the alphabet $2^{\{f_1, \ldots, f_k\}}$.
The DFA accepts the finite word $w=w_1 w_2 \ldots w_n$
if and only if the corresponding set of time-abstract paths in the
CTMC contributes to $\Pr_s(\phi)$, i.\,e.\@, it respects the order of the $f_i$.
The transformation does not require to construct the full DFA,
but only the product of the CTMC and the DFA.
We show that the product is a stratified CTMC,
and moreover, the measure $\Pr_s(\phi)$ is preserved.
This product can be constructed in linear time and space in the size of the CTMC and $k$.
Thus our method will be useful as the centerpiece of a full CSL model checker equipped with multiple until
 formulas.

  Recently, the decision algorithm by Aziz
  \emph{et al.\@} was shown to produce erroneous results on some non-stratified CTMCs~\cite{Jansen11-ErratumAziz}.
  Still, their algorithm is correct on stratified CTMCs.
  As an additional contribution, our measure-preservation theorem
   ensures the decidability of CSL model checking for general CTMCs.


\subsubsection*{Overview of the article.}
Section~\ref{sec:preliminaries} sets the ground for the paper.
In Section~\ref{sec:stratified_ctmcs} we introduce \emph{stratified CTMCs} formally.
The first main result is shown in Section~\ref{sec:product}:
it constructs a DFA for an until formula,
and then shows that the product is a stratified CTMC
and the relevant measures are preserved.
Section~\ref{sec:compute} discusses the computations in the product CTMC.
A model checking
algorithm is presented in Section~\ref{sec:model_checking}. Section~\ref{sec:related} discusses related
work, and the paper is concluded in Section~\ref{sec:conclusion}.

\section{Preliminaries}
\label{sec:preliminaries} 
This section presents the definition of Markov chains, probability
space, transient and steady-state distributions. For details please
refer to~\cite{Steward94,prakash,hermanns_ctmc}.
\subsection{Markov Chains}
\begin{defi}
A labeled \emph{discrete-time Markov chain} (DTMC) is a tuple $\dtmc$, where $S$
is a finite set of states, $\P:S\times S\to [0,1]$ is a
probability matrix satisfying $\sum_{s'\in
  S}\P(s,s')\in\{0,1\}$ for all $s\in S$, and
$L:S\to 2^{AP}$ is a labeling function.  

	A labeled \emph{continuous-time Markov chain} (CTMC) is a tuple $\ctmc$,
	where $S$ and $L$ are defined as for DTMCs,
	and $\R:S\times S\to \mathbb{R}_{\ge 0}$ is a rate matrix.
\end{defi}
For $A\subseteq S$, define $\R(s,A):=\sum_{s'\in A}\R(s,s')$, and let
$E(s):=\R(s,S)$ denote the \emph{exit rate} of $s$. A state $s$ is
called \emph{absorbing} if $E(s)=0$.
If $\R(s,s')>0$, we say that there is a transition from
$s$ to $s'$.
 
The transition probabilities in a CTMC are exponentially distributed
over time.
	If $s$ is the current state of the CTMC,
	the probability that some transition will be triggered within time $t$
	is $1-e^{-E(s) t}$.
	Furthermore, if $\R(s,s')>0$
	for more than one state $s'$,
	the probability to take a particular transition to $s'$
	is $\frac{\R(s,s')}{E(s)} \cdot \left( 1 - e^{-E(s) t} \right)$.
	The labeling function $L$ assigns to each state $s$
	the set of atomic propositions $L(s) \subseteq AP$
	which are valid in $s$.

	A CTMC $\C$ (and also a DTMC) is usually equipped
	with an initial state $s_\mathrm{init} \in S$
	or, more generally, an initial distribution
	$\alpha_\mathrm{init}:S\to [0,1]$
	satisfying $\sum_{s\in S}\alpha_\mathrm{init}(s)=1$.

\subsubsection*{Paths and probabilistic measures.}
A (sample) path is a right-continuous function
$\sigma:\mathbb{R}_{\ge 0} \to S$
(with the discrete topology on $S$).
Then, $\sigma(t)$ denotes the state occupied at time $t$.

For $i\in \mathbb{N}$, let
$\sigma_S[i]=s_i$ denote the $(i+1)$-th state visited, and $\sigma_T[i]=t_i$
denote the time spent in $\sigma_S[i]$.  For finite paths, $\sigma_T[n]$ is
defined to be $\infty$ if $\sigma_S[n]$ is the last (absorbing) state.
Let $\infpaths^\C$ denote the set of all (finite and infinite) paths,
and $\infpaths^\C(s)$ denote the subset of those paths starting from
$s$.

We sometimes use a different notation to describe a path,
namely a finite sequence $\sigma = \ps{s_0}{t_0}\psd{s_1}{t_1} s_n$
(meaning that $\sigma_S[i] = s_i$ and $\sigma_T[i] = t_i$ for all $i<n$,
and $\sigma_S[n] = s_n$ is an absorbing state),
or an infinite sequence $\sigma = \ps{s_0}{t_0}\psd{s_1}{t_1}$
if no absorbing state is hit.
The relation between the two notations is:
$\sigma(t) = s_i$
where $i$ is the smallest index with $t < \sum_{j=0}^i t_j$
(as remarked by \cite[p.~170]{prakash},
we have to use a strict inequality here for technical reasons,
not the non-strict inequality as in \cite{hermanns_ctmc}.).

Let $s_0,s_1,\ldots,s_k$ be states in $S$ with $\R(s_i,s_{i+1})>0$ for
all $0\le i < k$.  Let $I_0, I_1,\ldots, I_{k-1}$ be
nonempty intervals in $\mathbb{R}_{\ge 0}$.  The \emph{cylinder set}
$\mathit{Cyl}(s_0,I_0,\ldots, s_{k-1}, I_{k-1}, s_k)$ is defined by:
\[
\mathit{Cyl}(s_0,I_0,\ldots, s_{k-1}, I_{k-1}, s_k):=\{ \sigma \in\infpaths^{\C} \mid \forall 0\le i \le k.\
\sigma_S[i]=s_i\wedge \forall 0\le i <k.\ \sigma_T[i]\in I_i \}.
\]
Let $\F(\infpaths^\C)$ denote the smallest $\sigma$-algebra on
$\infpaths^\C$ containing all cylinder sets.
	For initial distribution $\alpha: S \to [0,1]$,
	a probability measure (denoted $\Pr_\alpha^\C$) on this $\sigma$-algebra
	is introduced as follows:
	$\Pr_\alpha^\C$ is the unique measure that satisfies:
	$\Pr^\C_\alpha(\mathit{Cyl}(s))$ equals $\alpha(s)$,
	and for $k>0$,
\[
\Pr^\C_\alpha(\mathit{Cyl}(s_0,I_0,\ldots,I_{k-1},s_k))
=\Pr^\C_\alpha(\mathit{Cyl}(s_0,I_0,\ldots,I_{k-2},s_{k-1}))
\cdot\tfrac{\R(s_{k-1},s_k)}{E(s_{k-1})}\cdot
\eta(I_{k-1})
\]
where $\eta(I_{k-1}):=\exp(-E(s_{k-1}) \inf I_{k-1}) - \exp(-E(s_{k-1}) \sup I_{k-1})$ 
is the probability to take a transition during time interval $I_{k-1}$.
(As a consequence, the probability of a cylinder set containing a point interval $[t,t]$ is $0$.)
If $\alpha(s)=1$ for some
state $s\in S$, we sometimes simply write $\Pr_s^\C$
instead of $\Pr_\alpha^\C$.  We omit the superscript
$\C$ if it is clear from the context.

\subsubsection*{Transient and steady-state probability.}
Starting with distribution $\alpha$, the transient probability vector
at time $t$, denoted by $\pi(\alpha,t)$, is the probability
distribution over states at time $t$. If $t=0$, we have
$\pi(\alpha,0)(s')=\alpha(s')$. For $t>0$, the transient
probability is given by: $\pi(\alpha,t) =
\pi(\alpha,0) e^{\mathbf{Q} t} $ where
$\mathbf{Q}:=\R-\mathit{Diag}(E)$ is the infinitesimal generator
matrix.  $\mathit{Diag}(E)$ denotes the diagonal matrix with
$\mathit{Diag}(E)(s,s)=E(s)$. The steady-state distribution is defined as the limit 
$\lim_{t\to\infty}\pi(\alpha,t)$, which always exists for finite CTMCs.

\subsection{Deterministic Finite Automata}

\begin{defi}
	A \emph{deterministic finite automaton} is a tuple
	$\B= (\Sigma, Q, q_{in}, \delta, F)$,
	where $\Sigma$ is a finite alphabet,
	$Q$ is a finite set of states,
	$q_{in}\in Q$ is an initial state,
	$\delta:Q\times\Sigma\partialto Q$ is a partial transition function,
	and $F\subseteq Q$ is a set of final states.
\end{defi}

	We call a finite sequence $w=w_1 w_2 \ldots w_n$ over
        $\Sigma$ a \emph{word} over $\Sigma$.  $w$ induces at most one
        path $\sigma(w)=\ps{q_0}{}\psd{q_1}{} q_n$ in $\B$
	where $q_0=q_{in}$
        and $q_{i}=\delta(q_{i-1},w_{i})$ for $i=1,\ldots,n$.  This
        word $w$, and also the corresponding path $\sigma(w)$, is
        \emph{accepting} if $\sigma(w)$ exists and $q_n \in F$.

\subsection{Continuous Stochastic Logic (CSL)}\label{section:syntax_csl}

We consider the branching-time temporal logic 
\emph{Continuous Stochastic Logic} (CSL) introduced by Aziz \emph{et
al.}~\cite{AzizSSB00},
which allows us to specify properties over CTMCs.
Its syntax is defined as follows:
\begin{align*}
  \Phi&:= a \mid\neg\Phi\mid\Phi\wedge\Phi\mid
  \Prob_{\trianglelefteq
    p}(\phi)\\
  \phi&:= \Phi_1 \until{I_1}\Phi_2 \until{I_2} \ldots \until{I_{k-1}} \Phi_k
\end{align*}
where $a\in AP$ is an atomic proposition,
$I_1, I_2,\ldots \subseteq \mathbb{R}_{\ge
  0}$ are nonempty left-closed intervals with rational bounds, $\mathord{\trianglelefteq} \in \{\mathord{<},\mathord{\le},\mathord{\ge},\mathord{>}\}$,
  $p \in \mathbb{Q} \cap [0,1]$, and $k\ge 2$.  We use the abbreviation
$\diam{I} \Phi=(\neg(a \wedge \neg a)) \until{I} \Phi$,
for an arbitrary atomic proposition $a$.
The syntax of CSL consists of
state formulas and path formulas: we use $\Phi,\Phi_1,\Psi,\Psi_1,\ldots$ for
state formulas and $\phi,\phi_1,\psi,\psi_1,\ldots$ for path formulas.

Let $\ctmc$ be a CTMC with $s\in S$.  The semantics
of most CSL state formulas is standard: $s\models a$ iff $a\in L(s)$;
$s\models \neg\Phi$ iff $s\not\models\Phi$; $s\models \Phi\wedge\Psi$
iff $s\models\Phi$ and $s\models \Psi$. For probabilistic formulas, we
have:
$$
s\models \Prob_{\trianglelefteq p}(\phi)
\mbox{ iff }\Pr_s\{\sigma\in \infpaths\mid\sigma\models\phi\}\trianglelefteq p
$$ where
$\Pr_s\{\sigma\in \infpaths\mid\sigma\models\phi\}$, or
$\Pr_s(\phi)$ for short, denotes the probability measure of the set of
all paths which start with $s$ and satisfy $\phi$.

The satisfaction relation for CSL path formulas is defined as follows:
let $\sigma$ be a path, and let $\phi= \Phi_1 \until{I_1}\Phi_2
\until{I_2} \ldots \Phi_k$ be a path formula.  Then $\sigma\models
\phi$ if and only if there exist real numbers $0\le t_1\le t_2 \le
\ldots \le t_{k-1}$ such that $\sigma(t_{k-1})\models\Phi_k$, and for
each integer $0<i<k$ we have $(t_i \in I_i) \wedge (\forall
t'\in [t_{i-1},t_i))(\sigma(t')\models \Phi_i)$, where $t_{0}$ is
defined to be $0$ for notational convenience.

	For a CSL path
formula $\phi=\Phi_1\until{[a_1,b_1)}\Phi_2 \until{[a_2,b_2)} \Phi_3$
with $a_2<a_1$, one can replace the second interval by $[a_1,b_2)$
	without changing the set of paths that satisfy the formula.
	Thus, we shall assume that the left endpoints
	-- and similarly, the right endpoints --
	of the intervals in multiple until formulas
	are always nondecreasing.

\section{Stratified CTMCs}\label{sec:stratified_ctmcs}
	The main challenge of model checking
	is the computation and the approximation of the probability $\Pr_s(\phi)$.
	We now introduce the class of \emph{stratified} CTMCs.
	This is the key
	for the computation of $\Pr_s(\phi)$.
	For now, the path formula $\phi$ contains pairwise different atomic propositions as subformulas.
	In Section~\ref{modelchecking-is-decidable},
	we shall see that this definition is easily generalized
	to formulas containing more complex subformulas.

Let $\ctmc$ be a CTMC. Let $\phi=f_1\until{I_1}f_2 \until{I_{2}}\ldots f_k$ be
a CSL path formula with pairwise different atomic propositions.
Moreover,  we let $F:=\{f_1,f_2,\ldots,f_k\}$, and 
$\sqsubseteq$ be an order on $F$ such that $f_i\sqsubseteq f_j$
iff $i\leq j$.
	For a state $s$, if the set $L(s) \cap F$ is not empty,
	we let $f_\mathrm{min}^s := \min_\sqsubseteq L(s) \cap F$ denote the least element $f_i$ with respect to the order $\sqsubseteq$.
  If such $f_j$ does
  not exist, we define $f_\mathrm{min}^s:=\bot$.
\begin{defi}[Stratified CTMC]\label{def:stra}
  We say that $\C$ is
  \emph{stratified} with respect to $\phi$ iff for all $s_1,s_2$, it
  holds that:
  \begin{iteMize}{$\bullet$}
  \item If  $f_\mathrm{min}^{s_1}=\bot$ or $f_\mathrm{min}^{s_1}=f_k$, then 
  $\mathbf{R}(s_1,s_2)=0$.
\item Otherwise (i.\,e.\@, $f_\mathrm{min}^{s_1}\neq \bot$ and $f_\mathrm{min}^{s_1}\neq f_k$), if $\mathbf{R}(s_1,s_2)>0$ and $f_\mathrm{min}^{s_2}\neq\bot$, then $f_\mathrm{min}^{s_1} \sqsubseteq
  f_\mathrm{min}^{s_2}$.
  \end{iteMize}
\end{defi}
%
%

A state $s$ with $f_\mathrm{min}^{s}=\bot$ is a bad state,
and a state with $f_\mathrm{min}^{s}=f_k$ is a good state.
	(Note that there may be other states satisfying $f_k$ as well.)
	Both good and bad states are absorbing.
	The intuition behind Def.~\ref{def:stra}
        is that paths reaching bad states will not satisfy $\phi$,
	while those reaching good states or other
	$f_k$-states may satisfy $\phi$
	(provided the timing constraints are also satisfied).
\begin{figure}[tbh]
\begin{center} 
\begin{picture}(55,31)(0,-8)
 \node[Nmarks=i](s0)(10,15){$s_0$}
  \node(s1)(10,0){$s_1$}
  \node(s2)(50,0){$s_2$} 
  \node(s3)(30,15){$s_3$}
  \node(s4)(50,15){$s_4$}
  \drawedge[ELside=r](s0,s1){$2$}
  \drawedge[curvedepth=2](s1,s2){$1$} 
  \drawedge[curvedepth=2](s2,s1){$1$} 
  \drawedge(s1,s3){$1$}
  \drawedge(s3,s2){$2$}
  \drawedge[ELside=r](s2,s4){$1$}
  \drawedge[ELside=r](s4,s3){$2$}
  \nodelabel[NLangle=90,ExtNL=y,NLdist=0.5](s0){$f_1,f_4$}
  \nodelabel[NLangle=270,ExtNL=y,NLdist=0.5](s1){$f_1,f_2,f_4$}
  \nodelabel[NLangle=270,ExtNL=y,NLdist=0.5](s2){$f_4$}
  \nodelabel[NLangle=90,ExtNL=y,NLdist=0.5](s3){$f_3$}
  \nodelabel[NLangle=90,ExtNL=y,NLdist=0.5](s4){$f_5$}
\end{picture} 
\end{center}
\caption{A non-stratified CTMC.\label{fig:ctmc_example}}
\end{figure}

\begin{exa}\label{example:phi1}
  Consider the path formula
 $\phi:=f_1\until{[0,2)}f_2\until{[2,4)} f_3 \until{[2,4)} f_4 \until{[3,5)}
  f_5$.
The CTMC in Fig.~\ref{fig:ctmc_example} is not stratified with respect to $\phi$:
we have
  $\R(s_2,s_1)>0$, however, $f^{s_2}_\mathrm{min} = f_4 \not\sqsubseteq
  f_1 = f^{s_1}_\mathrm{min}$.
	Deleting this edge and the transition out of $s_4$
	would result in a stratified CTMC with respect to $\phi$.
	\qed
\end{exa}

The notion of stratified CMTCs is the key to an efficient
approximation algorithm. The essential idea is that we can reduce the
model checking problem to one on a similar, stratified CTMC
that preserves the relevant reachability probabilities.
Further, our notion of stratified CTMCs solves a semantical problem in \cite{AzizSSB00}:
please refer to Section \ref{sec:usefullness} for details.

\section{Product CTMC}{\label{sec:product}}
Given a CTMC and a CSL path formula $\phi$, in this section we
construct a stratified CTMC with respect to $\phi$ preserving the
probability to satisfy $\phi$.  We first construct a
deterministic finite automaton for $\phi$ in
Subsection~\ref{sec:construction}. Then, in Subsection~\ref{subsec:product}
we build a product CTMC with the desired property.

\subsection{Automaton for a CSL Formula}\label{sec:construction}
For a path formula $\phi = f_1\until{I_1}f_2 \until{I_{2}}\ldots f_k$,
we first construct a simple deterministic finite automaton (DFA)
that describes the required order
of $f_1$-, $f_2$-, \ldots, $f_k$-states.
\begin{defi}[Formula automaton]\label{def.formula.automaton}
  Let $\phi=f_1\until{I_1}f_2 \until{I_{2}}\ldots f_k$ be a CSL path
  formula with pairwise different atomic propositions.
  Then, the \emph{formula automaton} $\B_{\phi} =(\Sigma, Q,
  q_{in}, \delta, F)$ is defined by: $\Sigma=2^{\{f_1,\ldots,f_k\}}$,
  $Q=\{q_1,q_2,\ldots, q_{k-1},q_k,\bot\}$ with $q_{in}=q_1$ and
  $F=\{q_1,\ldots,q_k\}$.  For $a\in\Sigma$, the transition relation
  $\delta$ is defined as follows:
\begin{enumerate}[(1)]
\item $\delta(q_i,a) = q_j$ if  $i<k$; $i\le j$; $f_i,f_{i+1}\ldots f_{j-1}\not\in a$; and $f_j\in a$;
\item $\delta(q_i,a) = \bot$
	if $i < k$ and the above clause does not apply;
\item $\bot$ and $q_k$ are absorbing.
\end{enumerate}
\end{defi}
As states $\bot$ and $q_k$ have no outgoing transitions, $\delta$ is a partial transition function.
Thus formula automata are actually partial DFAs.
The words accepted by $\B_\phi$ are finite traces $w\in \Sigma^*$
that can be extended to a trace $ww'\in \Sigma^\omega$
that satisfies the time-abstract
formula of the form $f_1 \until{} f_2 \until{} \ldots \until{} f_k$.
 The constructed finite automaton $\B_\phi$ for this
  special class of formulas is deterministic, the number of states is
  linear in $k$.  The number of transitions is $(k-1)2^k$; however, as we
  will see later,
  the product can be constructed in time (and size)
  linear in the size of the CTMC
  and in $k$.

\begin{figure}[tbh]
\begin{center}
\begin{picture}(107,58)(-20,-29)
 \node[Nmarks=ir](s1)(-10,0){$q_1$}
  \node[Nmarks=r](s2)(30,0){$q_2$}
  \node(bot)(70,25){$\bot$}
 \node[Nmarks=r](s3)(70,0){$q_3$} 
 \node[Nmarks=r](s4)(70,-25){$q_4$}
  \drawedge(s1,s2){$\neg f_1 \mywedge f_2$}
  \drawedge[curvedepth=8,ELpos=45,ELdist=0](s1,s3){$\neg (f_1{\vee}f_2) \mywedge f_3$}
  \drawedge[curvedepth=-9,ELpos=35,ELside=r,ELdist=0](s1,s4){$\neg (f_1{\vee}f_2{\vee}f_3) \mywedge f_4$}
  \drawedge[curvedepth=9,ELpos=35,ELdist=0](s1,bot){$\neg (f_1{\vee}f_2{\vee}f_3{\vee}f_4)$}
  \drawedge(s2,s3){$\neg f_2 \mywedge f_3$}
  \drawedge[ELpos=45,ELside=r,ELdist=0](s2,s4){$ \neg (f_2{\vee}f_3) \mywedge f_4$}
  \drawedge[ELpos=45,ELdist=0](s2,bot){$ \neg (f_2{\vee}f_3{\vee}f_4)$}
  \drawedge(s3,s4){$\neg f_3\mywedge f_4$}
  \drawedge[ELside=r](s3,bot){$\neg (f_3{\vee}f_4)$}
  \drawloop[loopangle=250](s1){$f_1$}
  \drawloop[loopangle=250](s2){$f_2$}
  \drawloop[loopangle=0](s3){$f_3$}
\end{picture}
\end{center}
\caption{$\B_\phi$ for $\phi = f_1 \until{} f_2 \until{} f_3 \until{} f_4$\label{fig:buechi}}
\end{figure}

\begin{exa}
	In Fig.~\ref{fig:buechi} the formula automaton for $k=4$
	is illu\-strated.
	The initial state is $q_1$,
	final states are marked with a double circle.
	The trans\-ition labels indicate
	which subsets of $AP$ are acceptable.
	For example, we have
	$\delta(q_1,\{f_1\})=\delta(q_1,\{f_1,f_2\})=q_1$,
	as both sets satisfy $f_1$.  \qed
\end{exa}

\subsection{Product CTMC}\label{subsec:product}

\begin{defi}[Product CTMC]\label{def.product.CTMC}
  Let $\ctmc$ be a CTMC and $\phi=f_1\until{I_1}f_2 \until{I_{2}}\ldots f_k$
  a path formula with pairwise different atomic propositions.
  Let $\B_{\phi}$ be as constructed above.
 The product $\C\times
  \B_{\phi}$ is a CTMC $(S',\R', L')$ where:
\begin{enumerate}[(1)]
\item $S'= S\times Q$,
\item \label{def.product.CTMC.transition}
	$\R'((s,q_i),(s',q'))$ equals $\R(s,s')$
	if  $s\models_\C f_i \vee f_{i+1} \vee \cdots \vee f_{k-1}$ and
	$q'=\delta(q_i, L(s') \cap \{ f_1, \ldots, f_k\})$,
	and equals $0$ otherwise,
\item the labeling function is defined by: 
  \begin{iteMize}{$\bullet$}
  \item $L'(s,q_i)=L(s)\cap \{f_i,f_{i+1},\ldots,f_k\}$ for $1\le i\le k$,
  \item $L'(s,\bot)=\emptyset$.
  \end{iteMize}
\item	Given an initial distribution $\alpha:S \to [0,1]$ of $\C$,
	the initial distribution of the product $\alpha': S \times Q \to [0,1]$
	is defined by: $\alpha'(s,q)$ equals $\alpha(s)$ if $q = \delta(q_{in},L(s) \cap \{ f_1, \ldots, f_k \})$, and equals $0$ otherwise.
\end{enumerate}
\end{defi}
\noindent The product CTMC contains two kinds of absorbing states.
	In general, states $(s,q)$ with $s\not\models \bigvee_{i=1}^{k}f_i$
	are absorbing in the product,
	as well as states reached through a transition
	that does not follow the prescribed order of $f_i$.
	These two kinds of states can be considered bad states.
	On the other hand, good states of the form $(s,q_k)$ with $s\models f_k$ are also absorbing.
	The behavior after such an absorbing state is irrelevant
	for the probability to satisfy $\phi$.
\begin{exa}
  Consider the CTMC in Fig.~\ref{fig:ctmc_example}, and consider the
  path formula $\phi_1:=f_1\until{[0,2)}f_2\until{[0,2)}f_3\until{[0,2)}f_4\until{[0,2)}f_5$.
	The path
	$\sigma_1:=\ps{s_0}{}\ps{s_1}{}\ps{s_3}{}\ps{s_2}{} s_4 \ldots$
	does, if $s_4$ is reached before time 2, satisfy $\phi_1$;
	however, the path $\sigma_2:=\ps{s_0}{}\ps{s_1}{}\ps{s_2}{}\ps{s_1}{}\ps{s_3}{}\ps{s_2}{} s_4 \ldots$ does not.
	The product of this CTMC with $\B_{\phi_1}$ is the CTMC depicted on the left of Fig.~\ref{fig:product}, which is stratified with respect to $\phi_1$.
	State $(s_4,q_5)$ is a \emph{good state}
	-- paths reaching this state before time 2
	correspond to paths
	satisfying $\phi_1$ in Fig.~\ref{fig:ctmc_example} --,
	while $(s_3,\bot)$ is a \emph{bad state.}

  For the same CTMC in Fig.~\ref{fig:ctmc_example}, consider the path
  formula $\phi_2:=f_1\until{[1,3)} f_2 \until{[1,3)} f_3 \until{[1,3)} f_4$.  The product CTMC
  $\C\times\B_{\phi_2}$ is depicted on the right of Fig.~\ref{fig:product}.
  This product is stratified with respect to
  $\phi_2$.  The absorbing state $(s_2,q_4)$ is a
  \emph{good state.} \qed
\end{exa}

\begin{figure}[tbp]
\begin{center} 
\begin{picture}(75,31)(0,-8)
\node[Nmarks=i](s0)(10,15){$s_0q_1$}
  \node(s1)(10,0){$s_1q_1$}
  \node(s2)(50,0){$s_2q_4$} 
  \node(s3)(30,15){$s_3q_3$}
  \node(s4)(50,15){$s_4q_5$}
  \node(s5)(70,0){$s_1q_4$}
  \node(s6)(70,15){$s_3\bot$}
  \drawedge[ELside=r](s0,s1){$2$}
  \drawedge(s1,s2){$1$} 
  \drawedge[curvedepth=2](s2,s5){$1$} 
  \drawedge[curvedepth=2](s5,s2){$1$}
  \drawedge(s1,s3){$1$}
  \drawedge(s3,s2){$2$}
  \drawedge[ELside=r](s2,s4){$1$}
  \drawedge[ELside=r,ELpos=55](s5,s6){$1$}
  \nodelabel[NLangle=90,ExtNL=y,NLdist=0.5](s0){$f_1,f_4$}
  \nodelabel[NLangle=270,ExtNL=y,NLdist=0.5](s1){$f_1,f_2,f_4$}
  \nodelabel[NLangle=270,ExtNL=y,NLdist=0.5](s2){$f_4$}
  \nodelabel[NLangle=90,ExtNL=y,NLdist=0.5](s3){$f_3$}
  \nodelabel[NLangle=90,ExtNL=y,NLdist=0.5](s4){$f_5$}
  \nodelabel[NLangle=270,ExtNL=y,NLdist=0.5](s5){$f_4$}
  \nodelabel[NLangle=270,ExtNL=y,NLdist=0.5](s6){$ $}
\end{picture} 
\hspace{1.5cm}
\begin{picture}(55,31)(0,-8)
 \node[Nmarks=i](s0)(10,15){$s_0q_1$}
  \node(s1)(10,0){$s_1q_1$}
  \node(s2)(50,0){$s_2q_4$}
  \node(s3)(30,15){$s_3q_3$}
\drawedge[ELside=r](s0,s1){$2$}
  \drawedge[ELside=r](s1,s2){$1$} 
\drawedge(s1,s3){$1$}
  \drawedge(s3,s2){$2$}
 \nodelabel[NLangle=90,ExtNL=y,NLdist=0.5](s0){$f_1,f_4$}
  \nodelabel[NLangle=270,ExtNL=y,NLdist=0.5](s1){$f_1,f_2,f_4$}
  \nodelabel[NLangle=270,ExtNL=y,NLdist=0.5](s2){$f_4$}
  \nodelabel[NLangle=90,ExtNL=y,NLdist=0.5](s3){$f_3$}
\end{picture} 
\end{center}
\caption{The reachable part of the product CTMC $\C\times\B_{\phi_1}$ (left) and $\C\times\B_{\phi_2}$ (right).\label{fig:product}}
\end{figure}

For a CTMC $\mathcal{C} = (S, \mathbf{R}, L)$ and a state $s \in S$,
we use $\mathcal{C}|_{s} = (S', \mathbf{R}', L')$
to denote the \emph{sub-CTMC} reachable from $s$, i.\,e.,
$S' \subseteq S$ is the states reachable from $s$, $\mathbf{R}'$ and $L'$ are functions restricted to $S'\times S'$ and $S'$, respectively.

\begin{thm}[Measure-preservation theorem]{\label{theorem:main}}
  Let $\ctmc$ be a CTMC and $\phi=f_1\until{I_1}f_2 \until{I_{2}}\ldots f_k$
  a path formula.
	Let $\B_\phi$ denote the formula automaton.
	For $s\in S$, let $s_B=(s,\delta(q_\mathit{in},L(s) \cap \{ f_1, \ldots, f_k\}))$.
	Then:
	\begin{enumerate}[\em(1)]
	\item	$\C\times \B_\phi|_{s_B}$ is stratified with respect to $\phi$;
	\item	$\Pr_s^\C(\phi)
		= \Pr_{s_B}^{\smash[b]{\C\times \B_\phi|_{s_B}}} (\phi)
		=\Pr_{s_B}^{\smash[b]{\C\times \B_\phi}} (\phi)$.
	\end{enumerate}
\end{thm}
\proof
 We prove first that $\C\times \B_\phi|_{s_B}$ is stratified with respect to
  $\phi$. Consider a state $(s,q)$. By definition of the
  product CTMC, if $(s,q)\not\models_{\C\times \B_\phi} \bigvee_{i=1}^{k-1}f_i$, 
  then $s \not\models_\C \bigvee_{i=1}^{k-1}f_i$ or $q \in \{ q_k, \bot \}$,
  so state $(s,q)$ is absorbing
  and therefore trivially satisfies the stratification conditions.
  Now assume that $(s,q)\models_{\C \times \B_\phi} \bigvee_{i=1}^{k-1}f_i$,
  $q \not\in \{ q_k, \bot \}$, and moreover assume $(s',q')$ is a state with $\R'((s,q),(s',q'))>0$
  (with $\R'$ as in Def.~\ref{def.product.CTMC}).
  By the definition
  of the transitions of $\B_\phi$, we have $q'=\delta(q,L(s') \cap \{ f_1, \ldots, f_k\})$.
  Now assume $f_\mathrm{min}^{(s',q')}\neq\bot$: it remains to be shown
  that $f_\mathrm{min}^{(s,q)} \sqsubseteq f_\mathrm{min}^{(s',q')}$.
	Let $1\le x \le k$ such that $f_x = f_\mathrm{min}^{(s,q)}$,
	and let $1\le y \le k$ be such that $q=q_y$.  The
  indices $x'$ and $y'$ are defined similarly for $(s',q')$. By
  definition of transitions of $\B_\phi$ and product CTMC, it is
  routine to verify that $x= y$ and $x'=y'$. Moreover, in $\B_\phi$,
  $q'=\delta(q,L(s') \cap \{ f_1, \ldots, f_k\})$ implies that $y'\geq y$, which shows that $x\leq x'$,
  proving $f_\mathrm{min}^{(s,q)} \sqsubseteq f_\mathrm{min}^{(s',q')}$.

  Now we prove the second clause.  Obviously, states not reachable
  from $s_B$ can be safely removed, thus $ \Pr_{s_B}^{\smash[b]{\C\times \B_\phi|_{s_B}}}
  (\phi)=\Pr_{s_B}^{\smash[b]{\C\times \B_\phi}} (\phi) $.
  We next prove that
  $\Pr_s^\C(\phi) = \Pr_{s_B}^{\smash[b]{\C\times \B_\phi}} (\phi)$
  by showing that $\sigma \mapsto \sigma_B$
  (the canonical mapping from paths in $\C$ to paths in $\C \times \B_\phi$)
  preserves the standard probability measures
  between the probability spaces.
  To this end, it is enough to show
  that given a cylinder set $C_B$ over $\C \times \B_\phi$,
  its reverse image $C = \{ \sigma | \sigma_B \in C_B \}$ satisfies
  $\Pr_s^\C(C) = \Pr_{s_B}^{\smash[b]{\C\times \B_\phi}}(C_B)$.

	Stated briefly, we now show
	that paths in $\C$ and in $\C \times \B_\phi$
	correspond to each other
	because we only add some (bounded) information about the past to the states.

	Let us first describe the canonical mapping $\sigma \mapsto \sigma_B$.
	Assume given a path
	$\sigma = \ps{s_0}{t_0}\psd{s_1}{t_1}$
	in $\C$.
  The corresponding path in $\C \times \B_\phi$ is
	$\sigma_B = \ps{(s_0,q^0)}{t_0}\psd{(s_1,q^1)}{t_1}$,
  where
  $q^0 = \delta(q_{in},L(s_0) \cap \{ f_1, \ldots, f_k\})$
  and $q^{i+1} = \delta(q^i,L(s_{i+1}) \cap \{ f_1, \ldots, f_k\})$ for all $i\ge 1$,
  as long as the $(s_i,q^i)$ are not absorbing.
  However, if $(s_n,q^n)$ is absorbing for some $n$,
  then $\sigma_B$ is defined to be the finite path
	$\ps{(s_0,q^0)}{t_0}\psd{(s_1,q^1)}{t_1} (s_n,q^n)$,
  where $(s_n,q^n)$ is the first absorbing state encountered.
  Note that $\sigma \models_\C \phi$
  iff $\sigma_B \models_{\C \times \B_\phi} \phi$.

  Let $C_B = \mathit{Cyl}((s_0,q^0),I_0,(s_1,q^1),\ldots,(s_n,q^n))$ and $C$
  be as above.
  By definition of a cylinder set,
  $\R'((s_i,q^i), (s_{i+1},q^{i+1})) > 0$ for all $i < n$,
  therefore $(s_i,q^i)$ is not absorbing (for $i < n$)
  and $q^{i+1} = \delta(q^i,L(s_{i+1}) \cap \{ f_1, \ldots, f_k\})$.
	Now assume that some path
	$\sigma = \ps{s'_0}{t_0}\psd{s'_1}{t_1} \in C$;
  then it must hold that $s'_0 = s_0$, $t_0 \in I_0$, $s'_1 = s_1$, $t_1 \in I_1$, \ldots, and $s'_n = s_n$.
  Therefore, $C \subseteq \mathit{Cyl}(s_0,I_0, s_1, \ldots, s_n)$.
  On the other hand, for all paths $\sigma \in \mathit{Cyl}(s_0,I_0, s_1, \ldots, s_n)$,
  it is easy to prove that $\sigma_B \in C_B$.
  So, $C \supseteq \mathit{Cyl}(s_0,I_0, s_1, \ldots, s_n)$,
  and together, $C = \mathit{Cyl}(s_0,I_0, s_1, \ldots, s_n)$.
  It is now an easy calculation to verify that $\Pr_s^\C(C) = \Pr_{s_B}^{\smash[b]{\C \times \B_\phi}}(C_B)$.

 The reverse image of the set of $\C \times \B_\phi$-paths satisfying $\phi$
  is exactly the set of $\C$-paths satisfying $\phi$.
  Since these sets are measurable,
  both can be decomposed into countable unions
  of corresponding cylinder sets in $\C$ and $\C \times \B_\phi$,
  respectively.
  Thus, the theorem follows.  
\qed

\section{Characterizing the Probability $\Pr_\alpha(\phi)$}{\label{sec:compute}}
For  a path formula $\phi$, together with a stratified CTMC with respect to $\phi$, 
this section aims at a recursive characterization of  the probability $\Pr_\alpha(\phi)$
starting from an arbitrary initial distribution $\alpha$.  

We first introduce some notation.
For an interval $I$ and $0\le x$, we let $I\ominus x$
denote the set $\{t-x\mid t\in I \wedge t\ge x\}$. For example,
$[3,8) \ominus 5 = [0,3)$.
Then, for
$\phi=f_1\until{I_1}f_2 \until{I_{2}} \ldots f_k$ and $x < \sup I_1$, we
let $\phi\ominus x$ denote the formula $f_1 \until{I_1\ominus x} f_2
\until{I_2\ominus x} \ldots f_k$.
	For $1 \le j', j \le k$, define
	$f_{j' \ldots j}:= \bigvee_{i=j'}^{j} f_i$;
	for $1 \leq j < k$, define
	$\phi_j:=f_j \until{I_j} f_{j+1}\until{I_{j+1}}\ldots f_k$.
	As a degenerate case of $\Pr_s(\phi_j)$,
	let $\Pr_s(f_k) := 1$ if $s \models f_k$ and $0$ otherwise.
 For  $\Phi$, we denote by
$\C[\Phi]$ the CTMC obtained by $\C$ by making states satisfying $\Phi$
absorbing -- by cutting transitions
 out of all states satisfying $\Phi$.
 Moreover, let $\I_\Phi$ denote the indicator matrix
defined by: $\I_\Phi(s,s) = 1$ if $s \models \Phi$, and
$\I_\Phi(s,s')=0$ otherwise.

\subsection{Left-Closed Intervals}
For the moment, we restrict our attention to until formulas
where all timing constraints have the form $I_i = [a_i, b_i)$.
The following theorem characterizes the probability for this case:

	\begin{thm}\label{theorem:mc}
		Let $\phi=f_1\until{I_1}f_2 \until{I_{2}} \ldots f_k$
		be a CSL path formula with pairwise different atomic propositions,
		and assume all $I_i = [a_i, b_i)$ are left-closed.
		Let $\ctmc$ be a stratified CTMC with respect to $\phi$.
		We write the vector
		$(\Pr_{s}^\C(\psi))_{s\in S}$
		as $\Pr_{(\cdot)}^\C(\psi)$.
		\begin{enumerate}[\em(1)]
		\item	\label{theorem:mc:a1g0}
			Assume $0 < a_1$.
			Then,
			\begin{equation}\label{eq:transient1}
			\Pr_\alpha^\C(\phi)
			= \pi^{\C[\neg f_1]}(\alpha,a_1) \cdot \I_{f_1}
			\cdot \Pr_{(\cdot)}^\C(\phi\ominus a_1)
			\end{equation}
where $\pi^{\C[\neg f_{1}]}(\alpha,a_1)$ is the transient distribution at time $a_1$ in the CTMC $\C[\neg f_{1}]$.
		\item	\label{theorem:mc:I1neIk}
			Assume $0= a_1 = \ldots = a_{j-1} < a_j < b_1$
			for some $j \in \{ 2, \ldots, k-1 \}$.
			Then,
			\begin{equation} \label{eq:transient2}
			\Pr_\alpha^\C(\phi)
			= \pi^{\C[\neg f_{1 \ldots j}]}(\alpha,a_j)
			\cdot \I_{f_{1 \ldots j}}
			\cdot \Pr_{(\cdot)}^{\C} (\phi\ominus a_j)
			\end{equation}
		\item	\label{theorem:mc:b1leaj}
			Assume $0= a_1 = \ldots = a_{j-1} < b_1 \leq a_j$
			for some $j \in \{ 2, \ldots, k-1 \}$.
			Let $j' \leq j$ be the largest integer
			such that $b_1 \not\in I_{j'-1}$.
			Then,
			\begin{equation}\label{eq:transient3}
			\Pr_\alpha^\C(\phi)
			= \pi^{\C[\neg f_{1 \ldots j}]}(\alpha,b_1)
			\cdot \I_{f_{j' \ldots j}}
			\cdot \Pr_{(\cdot)}^{\C[\neg f_{j' \ldots k-1}]} (\phi_{j'} \ominus b_1)
			\end{equation}

		\item	\label{theorem:mc:jeqk}

			Assume $0= a_1 = \ldots = a_{k-1}$.
			Let $j' \leq k$ be the largest integer
			such that $b_1 \not\in I_{j'-1}$.
			Then,
			\begin{equation}\label{eq:transient4}
			\Pr_\alpha^\C(\phi)
			= \pi^{\C[f_k]}(\alpha,b_1)
			\cdot \I_{f_{j' \ldots k}}
			\cdot \Pr_{(\cdot)}^{\C[\neg f_{j' \ldots k-1}]} (\phi_{j'} \ominus b_1)
			\end{equation}
			If $b_1 = \infty$,
			we replace $\pi^{\C[f_k]}(\alpha,b_1)$ in this equation
			by the corresponding steady-state distribution.
		\end{enumerate}
	\end{thm}

\noindent The key idea of the theorem is a \emph{property-driven
          transient analysis}. In the first clause we have $a_1>0$,
        thus for any path $\sigma$ satisfying $\phi$ it must hold
        $\sigma(t)\models f_1$ for all $t\in [0,a_1)$. Thus, we make
        all states satisfying $\neg f_1$ absorbing, and compute the transient
        distribution $\pi^{\C[\neg f_1]}(\alpha,a_1)$. Furthermore, the
        multiplication with the matrix $\I_{f_1}$ removes the
        probabilities in states satisfying $\neg f_1$ -- thus
        resulting in a subdistribution. Starting with this
        subdistribution, the formula will also be reduced by duration
        $a_1$. In the other clauses, we
        consider the interval $[0,a_j)$ or $[0, b_1)$, which is the common
          prefix of the intervals $I_1,\ldots,I_{j-1}$.
	Thus during this time the formula $f_{1\ldots j}$ must be
        satisfied. Here the assumption of stratification is
        crucial: otherwise one might be able jump forward and back
        between states satisfying $f_1$ and $f_j$, which is
        illustrated in the following example.

\begin{figure}[ht]
\begin{center}
\begin{picture}(95,12)(-10,-8)
  \node[Nmarks=i](s0)(0,0){$s_0$}
  \node(s1)(20,0){$s_1$}
  \node(s2)(40,0){$s_2$}
  \node(s3)(60,0){$s_3$}
  \node(s4)(80,0){$s_4$}
  \drawedge(s0,s1){$2$}
  \drawedge(s1,s2){$2$}
  \drawedge(s2,s3){$2$}
  \drawedge(s3,s4){$2$}
  \nodelabel[NLangle=270,ExtNL=y,NLdist=0.5](s0){$f_1$}
  \nodelabel[NLangle=270,ExtNL=y,NLdist=0.5](s1){$f_2$}
  \nodelabel[NLangle=270,ExtNL=y,NLdist=0.5](s2){$f_1$}
  \nodelabel[NLangle=270,ExtNL=y,NLdist=0.5](s3){$f_2$}
  \nodelabel[NLangle=270,ExtNL=y,NLdist=0.5](s4){$f_3$}
\end{picture} 
\end{center}
\caption{A CTMC with $\Pr_{s_0}(f_1 \until{[0,1)} f_2 \until{[0,1)} f_3) = 0$.\label{fig:ctmc_simple2}}
\end{figure}

\begin{exa}\label{exa:ctmc2}
  Consider the CTMC depicted in Fig.~\ref{fig:ctmc_simple2} and
  consider the path formula $\phi=f_1 \until{[0,1)} f_2 \until{[0,1)}
  f_3$. Obviously the probability of the set of paths starting from $s_0$ satisfying
  $\phi$ is $0$. Since the CTMC is not stratified with respect
  to $\phi$, Thm.~\ref{theorem:mc} cannot be applied directly: the product
  shall be constructed first. In the product
  CTMC, no states labelled with $f_3$ will be reached, thus giving the
  probability $0$, as desired.
\end{exa}

\subsubsection*{Proof of Thm.~\ref{theorem:mc}}
  We start with Eqn.~\eqref{eq:transient1}.  Let $a_1$ and the other notation be as in the
  theorem. For $s'\in S$, define the event $Z_{\ref{theorem:mc:a1g0}}(s'):=\{\sigma \mid
  \sigma(a_1)=s' \wedge\forall t \in [0, a_1).\ \sigma(t)\models f_1\}$,
  consisting of paths which occupy state $s'$ at time $a_1$ and
  occupy $f_1$-states during the time interval $[0, a_1)$. Obviously,
  $\{\sigma\mid \sigma\models\phi\}\subseteq \bigcup_{s'\models f_1}
  Z_{\ref{theorem:mc:a1g0}}(s')$.\footnote{Strictly
	speaking, this does not hold always
	because there may be paths that enter an $(f_2 \wedge \neg f_1)$-state
	exactly at time $a_1$;
	however, such paths are contained in a (generalization of) cylinder sets like $\mathit{Cyl}(s, [a_1, a_1], \ldots)$, whose measure is $0$.}
  Fix first $\alpha_s$ as an initial distribution with
  $\alpha_s(s)=1$ and $s\models f_1$. By the law of total probability, we
  have:
	\pagebreak[0]
\begin{align*}
  \Pr_{s}^\C(\phi)
&
= \sum_{s'\models f_1} \Pr_{s}^\C(Z_{\ref{theorem:mc:a1g0}}(s')) \cdot \Pr_{s}^\C(\phi \mid Z_{\ref{theorem:mc:a1g0}}(s')) \\
&
= \sum_{s'\models f_1} \pi^{\C[\neg f_1]}(s,a_1)(s')\cdot 
\Pr_{s}^\C(\phi \mid Z_{\ref{theorem:mc:a1g0}}(s')) \\
\intertext{%
The latter equality follows from the definition of $Z_{\ref{theorem:mc:a1g0}}(s')$.
By the Markov property of CTMCs:
}
&
= \sum_{s'\models f_1} \pi^{\C[\neg f_1]}(s,a_1)(s') \cdot \Pr_{s'}^\C(\phi\ominus a_1) \\
& = \sum_{s' \in S} \pi^{\C[\neg f_1]}(s,a_1)(s') \cdot \mathbf{1}_{s' \models f_1} \cdot \Pr_{s'}^\C(\phi\ominus a_1)
\end{align*}
where $\mathbf{1}_{s' \models f_1}$ is $1$ if $s' \models f_1$ and $0$ otherwise.
	Note that $\Pr_\alpha^\C(\phi) =
		\sum_{s\in S}\alpha(s) \Pr_s^\C(\phi) =
		\sum_{s\models f_1}\alpha(s) \Pr_s^\C(\phi)$,
	thus Eqn.~\eqref{eq:transient1} follows.

	We now jump to the proof of Eqn.~\eqref{eq:transient3}.
	This proof is more involved,
	but follows the same lines.
	Define the event
	$Z_{\ref{theorem:mc:b1leaj}}(s') := \{ \sigma \mid \sigma(b_1) = s' \wedge
		\forall t \in [0, b_1).\linebreak[1] \sigma(t) \models f_{1 \ldots j}\}$.
	Again, $\{ \sigma \mid \sigma \models \phi \} \subseteq
	\bigcup_{s' \models f_{j' \ldots j}} Z_{\ref{theorem:mc:b1leaj}}(s')$,
	and again, fix $\alpha_s$ as an initial distribution
	with $\alpha_s(s)=1$
	and $s\models f_{1 \ldots j}$.
	We have:
	\begin{align*}
		\Pr_{s}^\C(\phi)
		& =
		\sum_{s'\models f_{j' \ldots j}} \Pr_{s}^\C(Z_{\ref{theorem:mc:b1leaj}}(s'))
		\cdot \Pr_{s}^\C(\phi \mid Z_{\ref{theorem:mc:b1leaj}}(s')) \\
		& = \sum_{s'\models f_{j' \ldots j}}
			\pi^{\C[\neg f_{1 \ldots j}]}(s,b_1)(s') \cdot
			\Pr_{s}^\C(\phi \mid Z_{\ref{theorem:mc:b1leaj}}(s'))
	\end{align*}
	where the latter equality follows from the definition of $Z_{\ref{theorem:mc:b1leaj}}(s')$.
	Now let $\sigma\in Z_{\ref{theorem:mc:b1leaj}}(s')$, thus
       $\sigma(b_1) = s'$, and $\sigma(t)\models f_{1 \ldots j}$ for
        all $0\le t<b_1$.  
Let          $\sigma'$ denote the suffix path defined by $\sigma'(x):=\sigma(x+b_1)$.

	Now, $\sigma \models \phi$ implies
		that at time $b_1$, $\sigma$ has reached
		a state in a stratum from $q_{j'}, \ldots, q_j$,
		so $\sigma'$ satisfies $\phi_{j'} \ominus b_1$.
		On the other hand, every path $\sigma \in Z_{\ref{theorem:mc:b1leaj}}(s')$
		whose corresponding $\sigma'$ satisfies $\phi_{j'} \ominus b_1$
		also satisfies $\phi$
		(because $\C$ is stratified).
		Again,
		$\Pr_{s}^\C(\phi \mid Z_{\ref{theorem:mc:b1leaj}}(s')) =\Pr_{s'}^\C(\phi_{j'} \ominus b_1)$,
		thus
		\begin{align*}
			\Pr_{s}^\C(\phi)
			& = \sum_{s'\models f_{j' \ldots j}}
				\pi^{\C[\neg f_{1 \ldots j}]}(s,b_1)(s')
				\cdot \Pr_{s'}^\C(\phi_{j'} \ominus b_1) \\
			& = \sum_{s' \in S}
				\pi^{\C[\neg f_{1 \ldots j}]}(s,b_1)(s')
				\cdot \mathbf{1}_{s'\models f_{j' \ldots j}}
				\cdot \Pr_{s'}^\C(\phi_{j'} \ominus b_1)
			\quad.
		\end{align*}
		However, $\C$ needs not be stratified w.\,r.\,t.\@ $\phi_{j'} \ominus b_1$,
		so to simplify the sub\-se\-quent calculations,
		we restratify it:
		$\C[\neg f_{j' \ldots k-1}]$ is stratified
		w.\,r.\,t.\@ $\phi_{j'} \ominus b_1$.
	Eqn.~\eqref{eq:transient3}
	for general initial distribution $\alpha$ follows
	as in the case of Eqn.~\eqref{eq:transient1}.

	The proof for Eqn.~\eqref{eq:transient2} is similar to the proof for Eqn.~\eqref{eq:transient3},
	except that $b_1$ has to be replaced by $a_j$
	and $j'$ by $1$.

	For Eqn.~\eqref{eq:transient4}, we can again make a similar proof.
	First assume that $j' = k$.
	In that case, the paths that have reached an $f_k$-state
	at any time in the interval $I_{k-1} = I_1$
	are exactly the paths that satisfy $\phi$.
	They have the same probability as the paths in $\C[f_k]$
	that are in an $f_k$-state exactly at time $b_1$.
	Therefore,
	\pagebreak[0]
	\begin{equation}
		\Pr_{s}^\C(\phi)
		= \sum_{s'\models f_k}
			\pi^{\C[f_k]}(s,b_1)(s')
		= \sum_{s' \in S}
			\pi^{\C[f_k]}(s,b_1)(s') \cdot
			\mathbf{1}_{s' \models f_k}
		\label{eq:reach-fk-early}
	\end{equation}
	With the usual assumption $f_{k \ldots k-1} = \mathit{false}$,
	the theorem follows immediately.

	If $j' < k$, besides the paths mentioned above,
	other paths satisfy $\phi$, namely paths
	that reach an $f_k$-state during the interval $I_{k-1} \setminus I_1 = [b_1, b_{k-1})$
	(and avoid $f_k$-states earlier).
	These are the paths that satisfy
	$(f_1 \wedge \neg f_k) \until{I_1} \ldots \until{I_{k-2}} (f_{k-1} \wedge \neg f_k) \until{I_{k-1} \setminus I_1} f_k$.
	Their probability is, according to Eqn.~\eqref{eq:transient3},
	\begin{equation*}
		\sum_{s' \in S}
		\pi^{\C[\neg (f_{1 \ldots k-1} \wedge \neg f_k)]}(s,b_1)(s')
			\cdot \mathbf{I}_{s' \models f_{j' \ldots k-1} \wedge \neg f_k}
			\cdot \Pr_{s'}^{\C[\neg f_{j' \ldots k-1}]} (\phi_{j'} \ominus b_1)
	\end{equation*}
	Note that 
	$\C[\neg (f_{1 \ldots k-1} \wedge \neg f_k)] = \C[f_k]$.
	Adding this term to Eqn.~\eqref{eq:reach-fk-early}
	produces the desired probability.

	We still have to prove Eqn.~\eqref{eq:transient4}
	for $b_1 = \infty$.
	In that case, all timing constraints are trivial
	($[a_i, b_i) = [0, \infty)$)
	and $j' = k$.
	Therefore, $\Pr_{s}^\C(\phi)$ is just the probability
	to reach an $f_k$-state eventually,
	which is exactly
	$\lim_{b_1 \to \infty} \pi^{\C[f_k]} (s,b_1) \mathbf{I}_{f_k} \Pr_{(\cdot)}^{\C[\neg f_{k \ldots k-1}]}(f_k)$.
\qed

\subsection{Closed Intervals}

In Thm. \ref{theorem:mc}, we have considered formula $\phi$ with
left-closed intervals. Now we discuss that a slight generalization of
it can be used to handle closed intervals. Thus, below we assume that $I_i=[a_i,b_i]$.

The proof of Thm.~\ref{theorem:mc} can be extended easily
to hold also for closed intervals.
Clause~\ref{theorem:mc:b1leaj} may lead to formulas
containing degenerate intervals $[0,0]$:
As $b_1 \in [a_1, b_1]$,
often $j' = 1$ in this clause.
(We have to assume, as an additional simplification of notation, $I_0 := \emptyset$.)
As a consequence, $\varphi_{j'} \ominus b_1 = f_1 \until{[0,0]} f_2 \until{I_2 \ominus b_1} \ldots f_k$.

Further, if the original $\varphi$ already contained a degenerate interval,
say $a_1 = b_1$, so $I_1 = \{ a_1 \}$,
applying Clause~\ref{theorem:mc:a1g0} will also lead to a formula containing $[0,0]$.
These situations can be handled by the following lemma:

\begin{lem}
  Let $\phi=f_1\until{I_1}f_2 \until{I_{2}} \ldots f_k$ be a CSL path
  formula.  Let $\ctmc$ be a stratified CTMC with respect to $\phi$.
  Moreover, assume $I_1 = \ldots = I_{j-1} = [0,0]$ for $2\le j \le k$.
  Then, $\Pr_s^\C(\phi) = \Pr_s^\C(f_{j} \until{I_{j}} \ldots f_k)$ for all $s\in S$.
\end{lem}
\proof
  Assume a path $\sigma$ satisfies $\phi$.  The degenerate intervals force
  $t_1=t_2=\ldots=t_{j-1}=0$, thus no conditions relating to
  $f_1,\ldots,f_{j-1}$ need to be checked.
\qed

\begin{figure}[htb]
\begin{center}
\begin{picture}(45,12)(-10,-8)
  \node[Nmarks=i](s0)(0,0){$s_0$}
  \node(s1)(30,0){$s_1$}
  \drawedge(s0,s1){$2$}
  \nodelabel[NLangle=270,ExtNL=y,NLdist=0.5](s0){$f_1,f_3$}
  \nodelabel[NLangle=270,ExtNL=y,NLdist=0.5](s1){$f_2$}
\end{picture} 
\end{center}
\caption{A CTMC with $\Pr_{s_0}(f_1 \until{[0,1]} f_2 \until{[1,2]} f_3) \not= \Pr_{s_0}(f_1 \until{[0,1)} f_2 \until{[1,2]} f_3)$.\label{fig:ctmc_simple}}
\end{figure}
\begin{exa}
  Consider the CTMC in Fig.~\ref{fig:ctmc_simple} and the path
  formula $\phi=f_1 \until{[0,1]} f_2 \until{[1,2]} f_3$.
	Then, $\Pr_{s_0}(\phi)$ is the probability
	to stay in $s_0$ for at least one time unit
	($\psi(0, E(s_0) \cdot 1)$ in the notation of Section~\ref{uniformization-algorithm} below),
	since we can choose $t_1 = t_2 = 1$ if $\sigma_T[0]\ge 1$.
	Applying Clause~\ref{theorem:mc:b1leaj} of Thm.~\ref{theorem:mc},
	we get $j = 2$, $j' = 1$ and
	$\Pr_{s_0}(\phi)
	= \pi^{\mathcal{C}[\neg f_{1 \ldots 2}]}(s_0, 1) \cdot \mathbf{I}_{f_{1 \ldots 2}}
	\cdot \Pr^{\mathcal{C}[\neg f_{1 \ldots 2}]}_{( \cdot )}(f_1 \until{[0,0]} f_2 \until{[0,1]} f_3)
	= \pi^\mathcal{C}(s_0, 1) \cdot \mathbf{I} \cdot \Pr^\mathcal{C}_{( \cdot )}(f_2 \until{[0,1]} f_3)
	= (e^{-2}, 0) \cdot (1, 0)^T
	= e^{-2}$,
	the correct value.
	(In~\cite{icalp},
	we defined $j'$ slightly differently,
	producing $j' = 2$ and consequently $\Pr_{s_0}(\phi) = 0$.
	Our earlier definition worked only for left-closed intervals.)
 \qed
\end{exa}

Below we apply the theorem to two formulas, and thereby get the well-known
result~\cite{hermanns_ctmc} for the case of binary until for the case $k=2$.
	As above, $\C$ is stratified and $\varphi = f_1 \until{I_1} f_2 \until{I_2} \ldots f_k$.
\begin{enumerate}[(1)]
\item \emph{Reachability probability.} Assume that
  $I_1=\ldots=I_{k-1}=[0,b]$. Then, it holds $\Pr_s(\phi)=\Pr_s(\diam{[0,b]}
  f_k)$,
	which is the probability to reach an $f_k$-state within time $b$.
\item \emph{Interval reachability.} Assume that
  $I_1=\ldots=I_{k-1}=[a,b]$ with $a<b$. Then, it holds
  $\Pr_s(\phi)=\sum_{s'\models f_1}\pi^{\C[\neg f_1]}(s,a)(s')\cdot \Pr_{s'}(\diam{[0,b-a]} f_k)$,
	which is the interval reachability probability
	of staying in $f_1$-states until time $a$
	and then moving to an $f_k$-state before time $b$ has passed.
\end{enumerate}

\subsection{Other  Intervals}
First, the following lemma states properties of the probabilities for
binary until with different interval types:

\begin{lem}[Closure of Intervals for Binary Until]\label{lemma:closure}
Let $s\in S$.
Assume given two nonempty intervals $I$, $J$ such that $\inf I = \inf J$ and $\sup I = \sup J$.
Then, it holds:
\begin{enumerate}[\em(1)]
\item	If\/ $0 \in I \Leftrightarrow 0 \in J$, then
	$s \models \Prob_{\trianglelefteq p}(\Phi \until{I} \Psi)$
	iff $s \models \Prob_{\trianglelefteq p}(\Phi \until{J} \Psi)$
	for $\mathord{\trianglelefteq} \in \{ \mathord{<}, \mathord{\leq}, \mathord{\geq}, \mathord{>} \}$.
\item	Otherwise, assume w.\,l.\,o.\,g.\@ $0 \in I$ and $0 \not\in J$, and assume $0<p<1$.
	Then,
	$s \models \Prob_{\trianglerighteq p}(\Phi \until{I} \Psi) \wedge \Phi$
	iff $s \models \Prob_{\trianglerighteq p}(\Phi \until{J} \Psi)$,
	for $\mathord{\trianglerighteq} \in \{ \mathord{\geq}, \mathord{>} \}$.
	Similarly,
	$s \models \Prob_{\trianglelefteq p}(\Phi \until{I} \Psi) \vee \neg \Phi$
	iff $s \models \Prob_{\trianglelefteq p}(\Phi \until{J} \Psi)$,
	for $\mathord{\trianglelefteq} \in \{ \mathord{<}, \mathord{\leq} \}$.
\end{enumerate}
\end{lem}

\noindent
The lemma follows immediately from the definition of the measure of
cylinder set.  To see why we have to treat the case $\inf I = 0$
separately (not distinguished in~\cite{hermanns_ctmc}), assume that
$\Phi=\Prob_{\leq 0.1}(f_2\until{(0,1]} f_1)$
	and consider the CTMC depicted in Fig.~\ref{fig:ctmc_simple}:
	obviously we have $s_0 \models \Phi$
	as $s_0\not\models f_2$.
	However, $s_0 \not\models \Prob_{\leq 0.1}(f_2\until{[0,1]} f_1)$
	as $s_0$ satisfies $f_1$ directly.
	The formula $\Phi$ is equivalent
	to $\Prob_{\leq 0.1}(f_2 \until{[0,1]} f_1) \vee \neg f_2$.

For until formulas with arbitrary multiplicity, we have discussed the
case that all of the intervals are left-closed or closed.
Other cases can be handled in a way similar to Lemma~\ref{lemma:closure}.
However, to avoid too many technicalities, we skip these details.

\section{Model Checking Algorithm}{\label{sec:model_checking}}
	Let $\ctmc$ be a CTMC, $s\in S$,
	and $\Phi$ be a CSL formula.
	The model checking problem is to check whether $s\models\Phi$.
In the following two sections, we
        discuss that the model checking problem is decidable and
        provide an efficient algorithm for approximate computation of $\Pr_s(\psi)$.

\subsection{Model Checking CSL is Decidable}\label{modelchecking-is-decidable}
	The standard algorithm to solve CTL-like model checking problems
	recursively computes the sets of states satisfying $\Psi$, denoted by $\Sat(\Psi)$,
	for all state subformulas $\Psi$ of $\Phi$.
	For CSL, the cases where $\Psi$ is an atomic proposition,
	a negation or a conjunction are given by:
	$\Sat(a)=\{s\in S\mid a\in L(s)\}$,
	$\Sat(\neg\Psi_1)=S\backslash \Sat(\Psi_1)$
	and $\Sat(\Psi_1\wedge\Psi_2)= \Sat(\Psi_1)\cap \Sat(\Psi_2)$.

	The case that $\Psi$ is the probabilistic operator
	is the challenging part.
	Let $\Psi = \Prob_{\trianglelefteq p}(\phi)$
	with $\phi=\Psi_1\until{I_1}\Psi_2 \until{I_2}\ldots \Psi_k$.
	By the semantics, checking $\Psi$ is equivalent
	to checking whether $\Pr_{s}(\phi)$ meets the bound $\trianglelefteq p$,
	i.\,e.\@, whether $\Pr_s(\phi)\trianglelefteq p$.
	Assume that the sets $\Sat(\Psi_i)$ have been calculated recursively.
	We replace $\Psi_1, \ldots, \Psi_k$ by fresh (pairwise different) atomic propositions
	$f_1,\ldots,f_k$
	and extend the label of state $s$ by $f_i$ if $s\in
        \Sat(\Psi_i)$.  The so obtained path formula is $\psi:=f_1
        \until{I_1} f_2\until{I_2}\ldots f_k$, and obviously we have
        $\Pr_s(\phi)=\Pr_s(\psi)$.  
	The steps needed  to characterize $\Pr_s(\psi)$ are:
        \begin{enumerate}[(i)]
        \item Construct the formula automaton $\B_\psi$.
        \item Build the product $\C\times\B_\psi$,
	which by Thm.~\ref{theorem:main} is a stratified CTMC
	w.\,r.\,t.\@ $\psi$.
      \item Apply Thm.~\ref{theorem:mc} repeatedly to compute $\Pr_s(\psi)$.
        \end{enumerate}
        Thus, the decidability for the probabilistic formula reduces
        to checking whether $\Pr_s(\psi)\trianglelefteq p$ holds true
        in the product CTMC.  After applying Thm.~\ref{theorem:mc} a
        finite number of times, we see that $\Pr_s(\psi)$ reduces to a
        product of transient probabilities.
	We can now follow the argumentation in \cite{AzizSSB00}:
	Although the calculations differ slightly,
	$\Pr_s(\psi)$ still is a finite sum $\sum_k \eta_k e^{\delta_k}$
	(with algebraic $\eta_k$ and $\delta_k$).
	For such an expression, \cite{AzizSSB00} proved
	that it can be decided whether it is $\trianglelefteq p$,
	for $p \in \mathbb{Q}$.
	Thus, we still have:

\begin{thm}[\cite{AzizSSB00}, Thm.~1]
  Model checking CSL is decidable.
\end{thm}

\subsection{Usefulness of Stratification}\label{sec:usefullness}
Our notion of stratified CTMCs solves a semantical problem
in~\cite{AzizSSB00}, which we recently pointed out
in~\cite{Jansen11-ErratumAziz}.  Very briefly, Aziz \emph{et
  al.}~\cite{AzizSSB00} gave an algorithm that did not use the $t_i$
(in the semantics of until formulas) explicitly, which led to
incorrect results for non-stratified CTMCs.

Consider the CTMC depicted in Fig.~\ref{fig:ctmc_simple2} and the
formula $\phi=f_1 \until{[0,1)} f_2 \until{[0,1)} f_3$ in Example
\ref{exa:ctmc2}.  For this example, the algorithm in \cite{AzizSSB00}
calculates the probability that a path satisfies, a.\,o.,\@ the
conditions: it stays in $f_1 \vee f_2$-states during time $[0,1)$,
thus giving a wrong result. This problem does not occur provided that
the CTMC is stratified.

\subsection{Efficient Algorithm for Approximating $\Pr_s(\psi)$}
\label{uniformization-algorithm}
	We first explain
	how to combine steps (i) and (ii)
	mentioned above,
	without having to construct the full automaton $\B_\psi$.
	Most parts of the construction of $\C \times \B_\phi$ depend on $\C$ only
	and do not require much information about $\B_\phi$.
	For example, for the state space,
	it is enough to generate $k$ copies of every state in $\C$,
	which requires time $\mathcal{O}(\lvert S\rvert k)$.
	When constructing the transitions
	according to Clause~\ref{def.product.CTMC.transition} of Def.~\ref{def.product.CTMC},
	one has to check $q' = \delta(q_i, L(s') \cap \{ f_1, \ldots, f_k\})$,
	but even this can be done without actually constructing $\B_\phi$
	by using the definition of $\delta$
	(Def.~\ref{def.formula.automaton}) directly.
	Therefore, the overall time complexity to find all transitions of $\C \times \B_\phi$ is
	$\lvert \mathbf{R} \rvert$ times the number of copies that its source state may have,
	i.\,e.\@, $\mathcal{O}(\lvert \mathbf{R} \rvert k)$, which is also the maximal total number of transitions.

	The usual numerical algorithm
	to compute the matrix exponential $e^{\mathbf{Q} t}$
	is based on \emph{uniformization}~\cite{Steward94}.
	This algorithm executes most calculations on the uniformized DTMC.
	For a CTMC, we say that $\lambda$ is a uniformization rate
	if $\lambda\ge \max_{s\in S} (E(s) - \R(s,s))$.

\begin{defi}
  Let $\ctmc$ be a CTMC. The uniformized DTMC of $\C$ with respect to
  the uniformization rate $\lambda$ is $\mathit{uni}(\C)=(S,\P,L)$ where
  $\P(s,s') = \R(s, s')/\lambda$ if $s \neq s'$ and
	$\P(s,s) = 1-\P(s,S\setminus \{s\})$.
\end{defi}
 
Let
$\P$ denote the transition matrix of the uniformized DTMC
$\mathit{uni}(\C)$, thus it holds that $\P=I+\mathbf{Q}/\lambda$ where
$I$ denotes the identity matrix. For $t>0$, then:
\begin{align}\label{eq:uniform}
  \pi(\alpha,t) 
    = \pi(\alpha,0) e^{(\P-I)\lambda t}
    =\pi(\alpha,0) e^{-\lambda t}\sum_{i=0}^\infty 
      \frac{(\lambda t)^i}{i!} \P^i 
    = \sum_{i=0}^\infty \psi(i,\lambda t)
\vec{v}(i)
\end{align}
In this formula, $\psi(i,\lambda t) = e^{-\lambda t} \cdot
\frac{(\lambda t)^i}{i!}$ denotes the $i$-th Poisson probability with
parameter $\lambda t$, i.\,e.,
the probability to see precisely $i$ transitions within time $t$.
The vector $\vec{v}(i)$ is the transient
probability of $\mathit{uni}(\C)$ after $i$ transitions, i.\,e.\@,
$\vec{v}(i)= \pi(\alpha,0) \P^i$.
The infinite sum is approximated,
by picking $\mathcal{O}(\sqrt{\lambda t})$ terms with large $\psi(i, \lambda t)$,
using the Fox--Glynn algorithm~\cite{FoxG88,2011-Jansen-UnderstandingFoxGlynn}.
To find the $\vec{v}(i)$ for
Eqn.~\eqref{eq:uniform}, one requires $\mathcal{O}(\lambda t)$
matrix--vector multiplications~\cite{hermanns_ctmc}.  The following lemma states the complexity
of our algorithm:
	\begin{lem}[Complexity]\label{lemma:complexity}
          Let $\lvert \mathbf{R} \rvert$ denote the number of transitions of $\C$ and
          $\lambda\in \mathbb{R}_{>0}$ the \emph{uniformization rate}
          satisfying $\lambda=\max_{s\in S} (E(s) - \R(s,s))$.  For each
          formula $\phi=f_1\until{I_1} f_2\until{I_2}\ldots f_k$, the
          probability $\Pr_{s_B}^{\smash[b]{\C\times
              \B_{\phi}}}(\phi)$ can be approximated:
		\begin{iteMize}{$\bullet$}
		\item	in time in $\mathcal{O}(\lvert \mathbf{R} \rvert k \cdot \lambda b)$
			if $b=\sup I_{k-1}$ is finite,
		\item	in time in $\mathcal{O}(\lvert \mathbf{R} \rvert k \cdot \lambda b + (\vert S \rvert k)^3)$
			if $\sup I_{k-1}$ is infinite,
			where $\lvert S \rvert$ is the number of states in $\C$
			and $b=\max \left( \{ \inf I_{k-1} \} \cup \{\sup I_{i}|1 \le i < k\} \setminus \{\infty\} \right)$.
		\end{iteMize}
		The space complexity is in $\mathcal{O}(\lvert \mathbf{R} \rvert k)$.
	\end{lem}
  \proof
Recall that the formula automaton $\B_{\phi}$ is deterministic, and
  the size of the product automaton is $\mathcal{O}(\lvert \mathbf{R} \rvert k)$ which is both linear
  in the size of the CTMC and the
  formula. This proves the space complexity. 

  For the time complexity assume first $b<\infty$ with $b=\sup
  I_{k-1}$. Applying Thm.~\ref{theorem:mc}, the probability
  $\Pr_{s_B}^{\smash[b]{\C\times \B_{\phi}}}(\phi)$ can be expressed as a
  sequence of transient probability analyses, which can be efficiently
  approximated by a sequence of uniformization analyses. The
  complexity of these analyses is linear in the size of the product
  automaton, and also linear in $\lambda b$.

  For the second case $\sup I_{k-1}=\infty$, by Thm.~\ref{theorem:mc}, a sequence of transient probability analyses is
  followed by one steady-state analysis,
  which can be done with Gaussian elimination for the equation systems $\pi\cdot \mathbf{Q}'=0$ and $\sum_{s\in S'}\pi(s)=1$, the complexity of which
  is $\mathcal{O}((|S|k)^3)$. Thus the complexity for this case follows.
\qed

Thus, with the notion of stratified CTMC, we achieve polynomial complexity.
Our algorithm therefore improves the work of~\cite{AzizSSB00},
where only multiple until formulas with suitable timing constraints can be checked polynomially.
In the worst case, \cite{AzizSSB00} has to decompose a CSL formula
into $\mathcal{O}((k-1)^{k-1})$ formulas with suitable timing,
	thus resulting in an overall time complexity of
	$\mathcal{O}(\lvert \R \rvert \cdot \lambda b (k-1)^k )$
	or $\mathcal{O}((\lvert \R \rvert k \cdot  \lambda b + (\lvert S \rvert k)^3)\cdot (k-1)^{k-1} )$,
	respectively.

\section{Related Work}\label{sec:related}
The logic CSL was first proposed in~\cite{AzizSSB96}, in which the model checking problem is
shown to be decidable. Our paper gives a practical solution: it shows
that the relevant probabilities can be approximated efficiently. For the case of binary until
path formula, Baier \emph{et al.}~\cite{hermanns_ctmc} have presented an
approximate algorithm for the model checking problem. Their method can
be considered a special case of our approach.  

	Baier \emph{et al.}~\cite{BaierCHKS07-asCSL} defined a logic asCSL
	that uses so-called programs as path formulas,
	i.\,e.\@ regular expressions over state formulas and actions.
	Programs can express multiple until formulas of the form $\phi_1 \until{[0,b)} \phi_2 \until{[0,b)} \cdots \until{[0,b)} \phi_k$
	because asCSL cannot restrict the duration of individual program phases.
	The model checking algorithm translates the program to an automaton
	almost equal to the one in Fig.~\ref{fig:buechi}.
	Our work generalizes the method to multiple until formulas with multiple time bounds.

More recently, Donatelli \emph{et al.}~\cite{DonatelliHS09} have extended
CSL such that  path properties can be expressed via a
deterministic timed automata (DTA) with a single clock.
Chen \emph{et al.}~\cite{ChenHKM09-lmcs} take this approach further
and consider DTA specifications with multiple clocks as well.

In principle, one can translate a multiple until formula to a DTA with a single clock.
Its basic structure would look similar to Fig.~\ref{fig:buechi},
but Donatelli's and Chen's DTAs also include all timing information
and would have a size in $\mathcal{O}(k^2)$ --
an example construction with $k=4$ is given in Appendix~\ref{sec:translate}.
To check whether a CTMC satisfies a DTA specification,
they build the product of the two,
apply the region construction,
and then solve a system of integral equations.
	Chen's method, applied directly to our specifications,
	would amount to a complexity in
	$\mathcal{O}(k^4 \lvert S \rvert \lambda c + k^9 \lvert S  \rvert^3)$,
	where $c$ is the largest difference between time constraints
	(roughly comparable to $b$ in Lem.~\ref{lemma:complexity}).
Note that our algorithm has only a complexity in $\mathcal{O}(\lvert \R \rvert k \cdot \lambda b)$ if $b=\sup I_{k-1}<\infty$
or $\mathcal{O}(\lvert \R \rvert k \cdot \lambda b + (\lvert S \rvert k)^3)$ otherwise.

\section{Conclusion}\label{sec:conclusion}
In this paper we have proposed an effective approximation algorithm
for CSL with a multiple until operator. We believe that it is the
centerpiece of a broadly applicable full CSL model checker.

The
technique we have developed in this paper can also be applied to a
subclass of PCTL$^*$ formulas. Let $\phi=f_1\until{I_1}f_2 \until{I_2}\ldots
f_k$ be a CSL path formula. As we have seen in the paper, in case of
$I_1=\ldots=I_{k-1}=[0,\infty)$, our multiple until formula
$f_1\until{I}f_2\until{I}\ldots f_k$ corresponds to the LTL formula
$f_1\until{}(f_2\until{}(\ldots(f_{k-1}\until{}f_k)\ldots))$. In
general, $\phi$ is similar to a \emph{step-bounded} LTL formula
$\phi=f_1\until{[i_1,j_1]}f_2\until{[i_2,j_2]}\ldots f_k$ with
$i_1,j_1,\ldots$ integers specifying the step bounds. 
Such step-bounded until LTL formulas can be first transformed into
nested \emph{next-state} formulas, for example we have:
$
f_1\until{[2,3]}f_2=f_1\wedge X (f_1\wedge
X(f_2\vee (f_1\wedge X(f_2))))
$.
 The approach we have established in this paper can be
adapted slightly to handle this kind of formulas in complexity
linear in $j_{k-1}$ (assuming $j_{k-1}<\infty$). 

We conclude the paper by noting the connection of our DFA-based
approach with the classical \emph{B\"uchi-automaton}-based LTL model
checking algorithm by Vardi and Wolper~\cite{VardiW86}. The LTL
formula $\phi$ is first transformed into a B\"uchi automaton -- of
exponential size in the worst case -- accepting exactly the words
satisfying $\phi$. Then, model checking LTL can be reduced to
automata-theoretic questions in the product. Instead of B\"uchi
automata accepting infinite runs, we only need DFAs, which is due to
the simple form of the multiple until formula: it does not
encompass the full expressivity of LTL. This simplification, moreover, allows
us to get a DFA whose number of states is only linear in the length of
the CSL formula, and the size of the product automaton is then linear
in both the size of the CTMC and the length of the CSL formula.

\section*{Acknowledgement}
Lijun Zhang and Flemming Nielson are partially supported by IDEA4CPS
and MT-LAB (a VKR Centre of Excellence).  David N. Jansen and Holger
Hermanns are partially supported by DFG/NWO Bilateral Research
Programme ROCKS and by the European Union Seventh Framework Programme
under grant agreement no.\@ ICT-214755 (QUASIMODO). The work of Holger
Hermanns has received support by the European Union Seventh Framework
Programme under grant agreement no.\@ 295261 (MEALS). We thank Yang
Gao and Ming Xu for finding a bug in Thm.~\ref{theorem:mc} in an
early version of this paper.

\bibliographystyle{abbrv}
\bibliography{icalp11_260_zhang}

\appendix

\section{Translating Fig.~\protect\ref{fig:buechi} to a DTA for CSL\textsuperscript{TA}}\label{sec:translate}

\begin{figure}[tbp]
\begin{center}
\begin{picture}(120,113)(0,-6)
	\node[Nmarks=i,Nadjust=w](q10)(10,100){$q_{1,[0,a_1)}$}
	\nodelabel[NLangle=270,ExtNL=y,NLdist=0.3](q10){$f_1$}
	\node[Nadjust=w](q11)(40,100){$q_{1,[a_1,a_2)}$}
	\nodelabel[NLangle=270,ExtNL=y,NLdist=0.3](q11){$f_1$}
	\node[Nadjust=w](q12)(80,100){$q_{1,[a_2,a_3)}$}
	\nodelabel[NLangle=270,ExtNL=y,NLdist=0.3](q12){$f_1$}
	\node[Nadjust=w](q13)(120,100){$q_{1,[a_3,b_1)}$}
	\nodelabel[NLangle=320,ExtNL=y,NLdist=0.3](q13){$f_1 \!\!\wedge\!\! \neg f_4$}
	\node[Nadjust=w](q2t1)(10,78){$q'_{2,[a_1,a_2)}$}
	\nodelabel[NLangle=100,ExtNL=y,NLdist=0.3](q2t1){$f_2 \!\!\wedge\!\! \neg f_1$}
	\node[Nadjust=w](q2t2)(50,78){$q'_{2,[a_2,a_3)}$}
	\nodelabel[NLangle=100,ExtNL=y,NLdist=0.3](q2t2){$f_2 \!\!\wedge\!\! \neg f_1$}
	\node[Nadjust=w](q2t3)(90,78){$q'_{2,[a_3,b_2)}$}
	\nodelabel[NLangle=135,ExtNL=y,NLdist=0.3](q2t3){$f_2 \!\!\wedge\!\! \neg (f_1 \!\!\vee\!\! f_4)$}
	\node[Nadjust=w](q21)(10,50){$q_{2,[a_1,a_2)}$}
	\nodelabel[NLangle=270,ExtNL=y,NLdist=0.3](q21){$f_2$}
	\node[Nadjust=w](q22)(40,50){$q_{2,[a_2,a_3)}$}
	\nodelabel[NLangle=105,ExtNL=y,NLdist=0.3](q22){$f_2$}
	\node[Nadjust=w](q23)(80,50){$q_{2,[a_3,b_2)}$}
	\nodelabel[NLangle=165,ExtNL=y,NLdist=0.3](q23){$f_2 \!\!\wedge\!\! \neg f_4$}
	\node[Nadjust=w](q3t2)(50,22){$q'_{3,[a_2,a_3)}$}
	\nodelabel[NLangle=270,ExtNL=y,NLdist=0.3](q3t2){$f_3 \!\!\wedge\!\! \neg f_2$}
	\node[Nadjust=w](q3t3)(90,22){$q'_{3,[a_3,b_3)}$}
	\nodelabel[NLangle=235,ExtNL=y,NLdist=0.3](q3t3){$f_3 \!\!\wedge\!\! \neg (f_2 \!\!\vee\!\! f_4)$}
	\node[Nadjust=w](q32)(80,0){$q_{3,[a_2,a_3)}$}
	\nodelabel[NLangle=180,ExtNL=y,NLdist=0.3](q32){$f_3$}
	\node[Nadjust=w](q33)(120,0){$q_{3,[a_3,b_3)}$}
	\nodelabel[NLangle=40,ExtNL=y,NLdist=0.3](q33){$f_3 \!\!\wedge\!\! \neg f_4$}
	\node[Nmarks=r](q4)(120,50){$q_4$}
	\nodelabel[NLangle=0,ExtNL=y,NLdist=0.3](q4){$f_4$}
	\drawedge[ELpos=47](q10,q11){$x=a_1$}
	\drawedge(q11,q12){$x=a_2$}
	\drawedge(q12,q13){$x=a_3$}
	\drawedge[curvedepth=4,ELpos=25,ELdist=0](q12,q4){$x=a_3$}
	\drawedge[ELpos=60,ELdist=0](q2t1,q22){$x=a_2$}
	\drawedge[ELpos=60,ELdist=0](q2t2,q23){$x=a_3$}
	\drawedge[ELpos=35,ELdist=0](q2t2,q4){$x=a_3$}
	\drawedge[ELside=r,ELpos=47](q21,q22){$x=a_2$}
	\drawedge[ELside=r](q22,q23){$x=a_3$}
	\drawedge[curvedepth=8,ELpos=25,ELdist=0](q22,q4){$x=a_3$}
	\drawedge[ELpos=25,ELdist=0](q3t2,q33){$x=a_3$}
	\drawedge[ELpos=30,ELdist=0](q3t2,q4){$x=a_3$}
	\drawedge[ELside=r](q32,q33){$x=a_3$}
	\drawedge[curvedepth=-4,ELside=r,ELpos=18,ELdist=0](q32,q4){$x=a_3$}
	%
	\drawedge[ELpos=25,ELdist=0](q13,q23){$x=b_1$}
	%
	\drawedge[ELpos=38,ELside=r,ELdist=0](q2t3,q33){$x=b_2$}
	\drawedge[ELpos=34,ELdist=0](q23,q33){$x=b_2$}
	\drawloop[dash={1}{0.5},loopangle=90](q10){}
	\drawloop[dash={1}{0.5},loopangle=90](q11){}
	\drawloop[dash={1}{0.5},loopangle=90](q12){}
	\drawloop[dash={1}{0.5},loopangle=90](q13){}
	\drawloop[dash={1}{0.5},loopangle=180](q21){}
	\drawloop[dash={1}{0.5},loopangle=245](q22){}
	\drawloop[dash={1}{0.5},loopangle=245](q23){}
	\drawloop[dash={1}{0.5},loopangle=270](q32){}
	\drawloop[dash={1}{0.5},loopangle=270](q33){}
	\drawedge[dash={1}{0.5}](q11,q2t1){}
	\drawedge[dash={1}{0.5}](q12,q2t2){}
	\drawedge[dash={1}{0.5}](q13,q2t3){}
	\drawedge[dash={1}{0.5}](q2t1,q21){}
	\drawedge[dash={1}{0.5}](q2t2,q22){}
	\drawedge[dash={1}{0.5}](q2t3,q23){}
	\drawedge[dash={1}{0.5}](q2t2,q3t2){}
	\drawedge[dash={1}{0.5}](q2t3,q3t3){}
	\drawedge[dash={1}{0.5}](q22,q3t2){}
	\drawedge[dash={1}{0.5}](q23,q3t3){}
	\drawedge[dash={1}{0.5}](q3t2,q32){}
	\drawedge[dash={1}{0.5}](q3t3,q33){}
	\drawedge[dash={1}{0.5}](q13,q4){}
	\drawedge[dash={1}{0.5}](q2t3,q4){}
	\drawedge[dash={1}{0.5}](q23,q4){}
	\drawedge[dash={1}{0.5},ELpos=69,ELdist=0,ELside=r](q3t3,q4){$x<b_3$}
	\drawedge[dash={1}{0.5},ELside=r](q33,q4){$x<b_3$}
\end{picture}
\end{center}
\caption{\label{fig:CSLTA}
	A deterministic CSL\textsuperscript{TA}-timed automaton
	for
	$f_1 \until{[a_1,b_1)} f_2 \until{[a_2,b_2)} f_3 \until{[a_3,b_3)} f_4$.
}
\end{figure}

	As mentioned in Section~\ref{sec:related},
	Donatelli \textit{et al.}~\cite{DonatelliHS09} have extended CSL
	such that path properties can be expressed via a timed automaton.
In Fig.~\ref{fig:CSLTA}, we include a DTA
	corresponding to the formula
	$f_1 \until{[a_1,b_1)} f_2 \until{[a_2,b_2)} f_3 \until{[a_3,b_3)} f_4$
	with $0 < a_1 < a_2 < a_3 < b_1 < b_2 < b_3$.
	Dashed lines correspond to transition edges;
	solid lines to boundary edges.
The automaton has a single clock $x$.
The state label $q_{i,I}$ indicates
that time $t_{i-1}$ has passed
and that the current time is in the interval $I$.
	The guard $x < b_3$ is needed in states without boundary edge
	to ensure
	that $q_4$ is not entered too late.

	The automaton illustrates that CSL\textsuperscript{TA} may
        need a DTA with $\mathcal{O}(k^2)$ states, where $k$ is the
        number of phases in the multiple until-formula.

\end{document}